\documentclass[%
reprint,
superscriptaddress,
amsmath,amssymb,
aps,
prb,
floatfix
]{revtex4-2}

\usepackage{multirow}
\usepackage{graphicx}
\usepackage{dcolumn}
\usepackage{bm}
\usepackage[version=3]{mhchem}
\usepackage{nicefrac}

\usepackage{color}
\usepackage{xcolor}
\usepackage{dcolumn}
\usepackage{amssymb}
\usepackage{url}
\usepackage{hyperref}
\usepackage{xfrac}
\usepackage{tabularx}
\usepackage{xspace}
\usepackage{amsmath}
\usepackage[normalem]{ulem}
\usepackage{mathtools}
\usepackage{physics}
\usepackage{dsfont}
\usepackage{float}
\usepackage{braket}
\usepackage{placeins}

\def\be{\begin{equation}}
\def\ee{\end{equation}}
\def\bea{\begin{eqnarray}}
\def\eea{\end{eqnarray}}
\def\ba{\begin{array}}
\def\ea{\end{array}}

\begin{document}
\title{Symmetry breaking and competing valence bond states in the star lattice Heisenberg antiferromagnet}
\author{Pratyay Ghosh}
\email{pratyay.ghosh@epfl.ch}
\affiliation{Institute of Physics, Ecole Polytechnique Fédérale de Lausanne (EPFL), CH-1015 Lausanne, Switzerland}

\author{Jan Koziol}
\affiliation{Department Physik, Friedrich-Alexander-Universität Erlangen-Nürnberg (FAU), D-91058 Erlangen, Germany}

\author{Samuel Nyckees}
\affiliation{Institute of Physics, Ecole Polytechnique Fédérale de Lausanne (EPFL), CH-1015 Lausanne, Switzerland}

\author{Kai Phillip Schmidt}
\affiliation{Department Physik, Friedrich-Alexander-Universität Erlangen-Nürnberg (FAU), D-91058 Erlangen, Germany}

\author{Fr\'ed\'eric Mila}
\affiliation{Institute of Physics, Ecole Polytechnique Fédérale de Lausanne (EPFL), CH-1015 Lausanne, Switzerland}

\begin{abstract}
We investigate the ground state phase diagram of the spin-$1/2$ antiferromagnetic Heisenberg model on the star lattice using infinite projected entangled pair states (iPEPS) and high-order series expansions. The model includes two distinct couplings: $J_d$ on the dimer bonds and $J_t$ on the trimer bonds. While it is established that the system hosts a valence bond solid (VBS) phase for $J_d \ge J_t$, the ground state phase diagram for $J_d < J_t$ has remained unsettled. Our iPEPS simulations uncover a first-order phase transition at $J_d/J_t \approx 0.18$, significantly lower than previously reported estimates. Beyond this transition, we identify a close competition between two valence bond crystal (VBC) states: a columnar VBC and a $\sqrt{3} \times \sqrt{3}$ VBC, with the latter consistently exhibiting lower energy across all finite bond dimensions.
The high-order series expansion supports this by finding that the $\sqrt{3} \times \sqrt{3}$ VBC state indeed becomes energetically favorable, but only at sixth order in perturbation theory, revealing the subtle nature of the competition between candidate states.
\end{abstract}

\maketitle

\section{Introduction}
The complex interplay of geometric frustration and quantum fluctuations in low-dimensional quantum magnets is known to deliver exotic ground states without classical counterparts~\cite{frustrationbook,Diepbook}, exemplified by valence bond solids (VBS) and valence bond crystals (VBC)~\cite{BOT-Sachdev,AKLT,Verstraete2004a,Ghosh-Spin-1_Kagome,Ghosh2023a}~\footnote{We adopt the convention to distinguishing between a valence bond solid (VBS), which preserves translational symmetry, and a valence bond crystal (VBC), which breaks it.}, nematic states~\cite{PhysRevLett.96.027213,Iqbal2016}, and quantum spin liquids \cite{PhysRevLett.86.1881,Moessner2001,PhysRevB.65.165113}. The eleven Archimedean lattices~\cite{Grunbaum2016,Richter2004,pac2025magneticorderingoutofplaneartificial}, i.e., the two-dimensional (2D) lattices obtained from planar tessellations by regular polygons, where every vertex is topologically equivalent, are prominent in the study of quantum magnetism due to their abundance in real materials. These include examples of square~\cite{PhysRevLett.93.167202,RevModPhys.63.1,Bertinshaw2019-lq}, triangular~\cite{Gao2022-fm,PhysRevB.109.035110,PhysRevB.111.L180402}, honeycomb~\cite{PhysRevB.82.064412,PhysRevLett.108.127203,Ghosh_Hida_Model_of_Kagome}, kagome~\cite{Kagome_Comp1,Kagome_Comp2}, Shastry-Sutherland~\cite{Kayegama1999,Brunt2018-dl,yadav2025observationunprecedentedfractionalmagnetization}, and maple leaf~\cite{Fennell2011,Haraguchi2018,Ghosh2023b,Ghosh2023,Lake2025,Ghosh2025} manifestations. The quantum antiferromagnets (AFMs) on the  triangular lattice, and its derivatives obtained from site depletion, such as the maple-leaf lattice and the kagome lattice, are of particular interest due to their innate geometric frustrations. Although the spin-$1/2$ Heisenberg AFMs on the triangular and the maple-leaf lattices, coordination numbers of $6$ and $5$, respectively, exhibit magnetic ordering~\cite{PhysRevLett.133.176502,Farnell2011,Beck2024}, the stronger fluctuation effects for a lower coordination number of $4$, i.e., the kagome lattice, leads to a magnetically disordered ground state (GS)~\cite{PhysRevResearch.4.043019,Schnabel2012-ky}, a promising host for a quantum spin liquid~\cite{PhysRevLett.118.137202,Yan2011}.

\begin{figure*}[t]
    \includegraphics[width=0.95\textwidth]{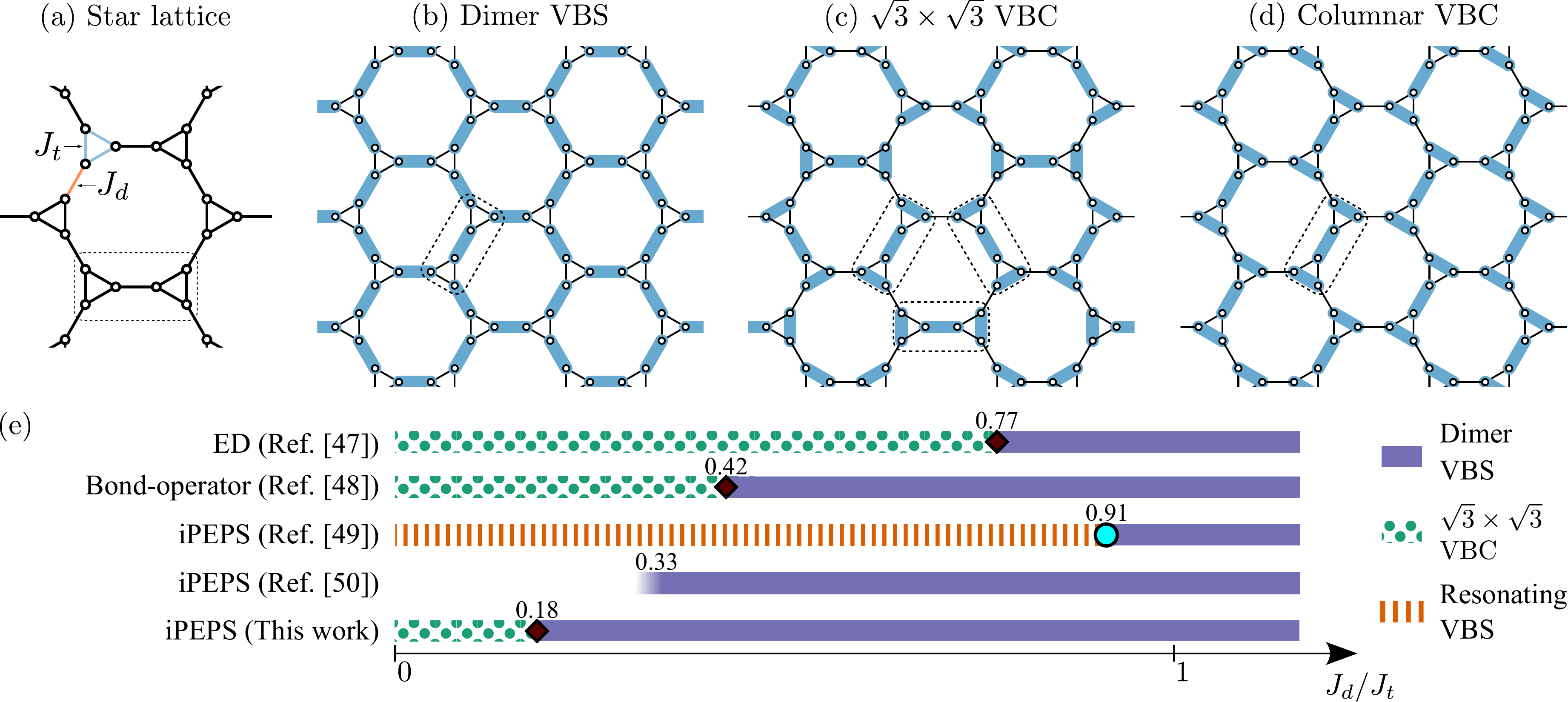}
    \caption{(a) The star lattice with its six-site unit cell, and the two symmetry-inequivalent nearest-neighbor bonds with Heisenberg exchange couplings $J_d$ (`dimer bonds') and $J_t$ (`trimer bonds'). (b)-(d) The candidate valence bond ground states of \eqref{eq-hamil} for $J_d,J_t>0$. The blue ellipses indicates strong singlet amplitudes (also throughout the article). (b) Fully-symmetric dimer VBS state. (c) $C_3$ symmetry-breaking VBC with $\sqrt{3}\times\sqrt{3}$ order. (d) $C_3$ breaking columnar VBC with six-site unit-cell. The $\sqrt{3}\times\sqrt{3}$ VBC and the columnar VBC are both three-fold degenerate. (e) Summary of previously reported phase diagrams for the system, alongside our result, as a function of $J_d/J_t$. ED predics a transition at $J_d/J_t \approx 0.77$~\cite{Star-ED}; Gutzwiller-projected wavefunction and bond-operator approaches suggest a first-order phase transition at $J_d/J_t \approx 0.42$~\cite{Star_PSG}; iPEPS studies report either a continuous transition at $J_d/J_t \approx 0.91$~\cite{Star-iPEPS1} or no transition at all down to $J_d/J_t \approx 0.33$~\cite{Star-iPEPS2}. The first two studies propose a $\sqrt{3} \times \sqrt{3}$ VBC as the favored state for $J_t \gg J_d$, while the iPEPS study Ref.~\cite{Star-iPEPS1} finds a resonating VBS with a six-site unit cell. Our result indicates a first-order transition out of the dimer VBS phase at $J_d/J_t = 0.185(5)$ into the $\sqrt{3} \times \sqrt{3}$ VBC.} \label{fig-vbs}
\end{figure*}

Among the Archimedean lattices with triangular motifs, the star lattice, shown in Fig.~\ref{fig-vbs} (a), has the lowest coordination number of $3$. This suggests an enhanced influence of quantum fluctuations, possibly leading the AFM on the lattice to realize physics richer than the kagome AFM. In addition, the star lattice inherently contains two symmetry-inequivalent nearest-neighbor bonds [see Fig.~\ref{fig-vbs} (a)]; varying the couplings of these bonds serves as the tunable parameter, facilitating phase transitions. Furthermore, there are iron acetate~\cite{Star_Fe} and hybrid copper sulfate~\cite{Star_Cu} that can realize star lattice spin systems, the latter of which realize no long-range magnetic order down to $0.1$~K~\cite{Star_Cu,PhysRevB.109.L180401}.

In this article, we study the spin-$1/2$ Heisenberg AFM on the star lattice described by the Hamiltonian
\be\label{eq-hamil}
\hat{H}=J_t\sum_{\langle ij\rangle \in t}\hat{\mathbf{S}}_i\cdot\hat{\mathbf{S}}_j+J_d\sum_{\langle ij\rangle \in d}\hat{\mathbf{S}}_i\cdot\hat{\mathbf{S}}_j. 
\ee 
The two terms represent the trimer bonds ($t$) and inter-triangle dimer-links ($d$), respectively. The ground-state phase diagram of the model has already been extensively studied~\cite{Star-ED2,Star-ED,Star_PSG,Star-iPEPS1,Star-iPEPS2}. It is established that the GS for $J_t=J_d$ is magnetically disordered, making the star lattice Heisenberg antiferromagnet (SLHA) the only other spin-$1/2$ Heisenberg model on an Archimedean lattice, besides the kagome, without a magnetic order. This state is a VBS with strong singlets on the $J_d$ bonds, deemed as \emph{dimer VBS} [see Fig.~\ref{fig-vbs} (b)]. Naturally, this VBS is also stable for $J_d>J_t$. In contrast, questions regarding the GS phase diagram for $J_d<J_t$ remain unsettled. Exact diagonalization (ED) suggests a phase transition out of the dimer VBS phase at $J_d/J_t\approx0.77$~\cite{Star-ED}. A subsequent work using Gutzwiller projected states and a bond-operator mean-field approach reports this critical ratio, a level crossing, to be $J_d/J_t\approx0.42$~\cite{Star_PSG}. Infinite projected entangled pair states (iPEPS) study by Jahromi and Or\'us~\cite{Star-iPEPS1}, however, identifies a continuous phase transition to occur at $J_d/J_t\approx0.91$, whereas another iPEPS study by Ran et. al.~\cite{Star-iPEPS2} reports no phase transition out of the dimer VBS phase for $J_d/J_t\in[0.33,\infty)$. These results (along with our own findings) are summarized in Fig.~\ref{fig-vbs} (e) for direct comparison. 

The nature of the GS beyond the transition is also under debate. Refs.~\cite{Star-ED,Star_PSG} report a VBC state exhibiting $\sqrt{3}\times\sqrt{3}$ order [shown in Fig.~\ref{fig-vbs} (c)] to be the likely candidate. It can be argued that the GS for $J_t\gg J_d$ should be a singlet covering which maximizes the number of occupied $J_t$ bonds; the $\sqrt{3}\times\sqrt{3}$ VBC is a member of this manifold. The iPEPS results in Ref.~\cite{Star-iPEPS1}, on the contrary, finds a resonating VBS GS with a six-site unit-cell, yielding an energy per site of $-0.255155$ (for bond dimension $D = 9$) at coupling parameters $(J_d, J_t) = (0.05, 1.0)$. A closer look reveals that this state cannot be the GS for $J_t\gg J_d$: The \mbox{$\sqrt{3}\times\sqrt{3}$} VBC is constructed with ``bowtie" blocks
\begin{center}
  \includegraphics[width=0.2\columnwidth]{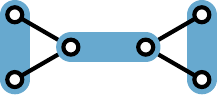}
\end{center}
with energy density of $-J_t/4-J_d/8$. The rest of the interactions that connect these blocks introduce no correction to the energy in first-order perturbation theory and the second-order will only lower the energy further. We can, thus, estimate an upper bound of the energy of the $\sqrt{3}\times\sqrt{3}$ VBC at $(J_d,J_t)=(0.05,1.0)$ to be $-0.25625$, which rules out the resonating VBS as a GS, leaving out the $\sqrt{3}\times\sqrt{3}$ VBC as the only proposed candidate ground state.

\begin{figure*}[t]
    \includegraphics[width=0.9\textwidth]{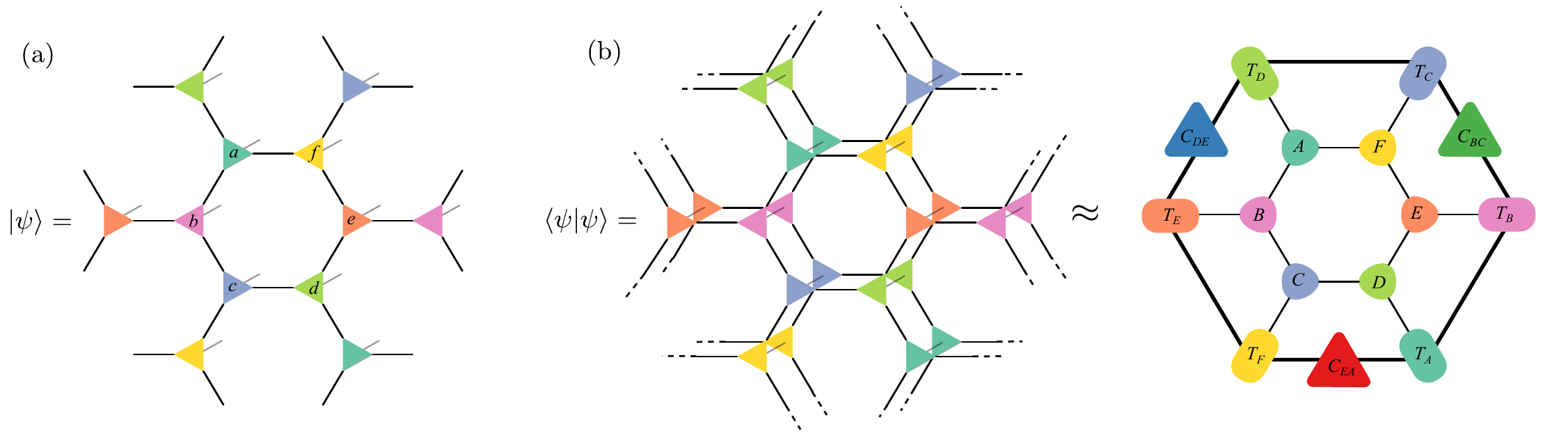}
    \caption{(a) iPEPS representation of a 2D wave function on the honeycomb lattice with a six-site unit cell. The black lines represent the virtual legs of dimension $D$ while the gray lines represent the physical legs of dimension $d=8$. (b) The overlap of wavefunctions is represented as the contraction of an infinite two-dimensional tensor network. The overlap of the tensors $a^{[\mathbf{x}]\dagger}a^{[\mathbf{x}]}$ is mapped to a local tensor $A^{[\mathbf{x}]}$ with dimensions $D^2 \times D^2 \times D^2$. This contraction is approximated using an environment composed of row and corner tensors. The bold black lines indicate the environment bond dimension $\chi$.} \label{fig-TN}
\end{figure*}

Motivated by this observation, we revisit the problem with tensor network methods, in particular iPEPS \cite{Jiang2008,Jordan2008,Verstraete2004a,Verstraete2004b,Gu2008} to capture the nature of the ground state beyond the dimer VBS phase directly in the thermodynamic limit. iPEPS is a variational ansatz where the wavefunction on an infinite 2D system is represented by a tensor network. The accuracy of the ansatz can be systematically controlled by the bond dimension $D$. This approach has successfully described various frustrated spin models, as well as bosonic and fermionic systems \cite{Jordan2008,PhysRevB.84.041108,Corboz2014,PhysRevB.85.125116,PhysRevB.88.155112,PhysRevB.94.075143}. We find the dimer VBS GS for $J_d=J_t$ in line with the previous studies. However, we find this state to be remarkably stable and a phase transition out of this phase only occurs at $J_d/J_t=0.185(5)$. Beyond this phase transition, we observe two states that closely compete for the ground state; one is the $\sqrt{3}\times\sqrt{3}$ VBC state and the other is a $C_3$ breaking VBC with six-site unit-cell.
We call this new candidate VBC GS a \emph{columnar VBC} due to its arrangement [see Fig.~\ref{fig-vbs} (d)]. For all finite $D\le11$, the columnar VBC consistently has higher energy than the $\sqrt{3}\times\sqrt{3}$ VBC; though the difference is only $\sim10^{-5}$. Unless otherwise stated, we set $J_t=1$ as the energy scale. As the $1/D$ scaling used to estimate the infinite-$D$ energy is empirical, the conclusion that the $\sqrt{3}\times\sqrt{3}$ VBC is the ground state in the infinite system—based solely on iPEPS calculations—should be treated with caution. As they are both made from the same blocks there is no simple argument, perturbative or otherwise, to rule one out. Therefore, we resort to a more involved perturbative analysis which not only confirms that $\sqrt{3}\times\sqrt{3}$ VBC to be the ground state of the system, but also offers a better understanding of the splitting between the competing states. The perturbative analysis is performed by deforming the Hamiltonian into the limit of uncoupled bowtie blocks and treating the couplings between the bowties as a perturbation. 
On a technical level, we perform a full graph-based perturbative linked-cluster expansion \cite{Gelfand2000,Oitmaa2006} for the non-degenerate energies of the two states in question using the L\"owdin partitioning formalism \cite{Loewdin1962,Yao2000,Kalis2012}.
We calculate the series in order seven in the inter-bowtie coupling. 
By construction, the series expansion technique provides the quantities directly in the thermodynamic limit for a given order.
Quite extraordinarily, we see that the energies of the two states remain identical up to fifth-order in perturbation. 
The graphs which splits the energy between the \( \sqrt{3} \times \sqrt{3} \) and columnar VBC states first appear at sixth order which is consistent with the observation of a splitting of the order of $10^{-5}$ in the iPEPS calculations.
Furthermore, We see that this splitting becomes more prominent with increasing order.

The remaining parts of the article are organized as follows. In Sec.~\ref{sec:mehods}, we first discuss the details of the iPEPS methodology we use for evaluating the GS of the system, followed by the details of the high-order series expansion. Thereafter, in Sec.~\ref{sec:results}, we report our results, investigate the phase transition, and the nature of the GS for $J_t\gg J_d$. Finally, in Sec.~\ref{sec:conclusions}, we conclude and put forward outlooks for future studies.

\section{Methods}\label{sec:mehods}
\subsection{Infinite projected entangled pair states (iPEPS)}
The iPEPS ansatz represents the GS wavefunction of a 2D system as an infinite 2D tensor network [Fig.~\ref{fig-TN} (a)]. While it is traditionally formulated on the square lattice, it can naturally be extended to other lattice geometries. In the iPEPS approaches, individual spin-$1/2$'s can be represented as tensors of physical dimension $d=2$; the approach is often generalized to use tensors of larger physical dimensions which represent a dimer or plaquette constituted of multiple spin-$1/2$'s \cite{Corboz2014,Corboz_SSM}. In this work, we consider a 2D tensor network where each $J_t$-trimer of the star lattice is represented by a tensor
\begin{center}
   \includegraphics[width=0.5\columnwidth]{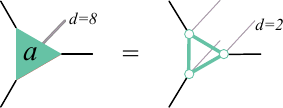}
\end{center}
with a physical leg of dimension $d=2^3$. This, in turn, leads to the representation of the GS as a tensor network defined on the honeycomb lattice with tensors $a_{i,j,k,s}^{[\mathbf{x}]}$ [see Fig.~\ref{fig-TN} (a)], where the indices $i,j,k$ refer to the virtual space with dimension $D$, while the index $s$ represents the physical space. The position of each tensor within the unit-cell is labeled by $[\mathbf{x}]$. 

\begin{figure*}[t]
    \includegraphics[width=0.95\textwidth]{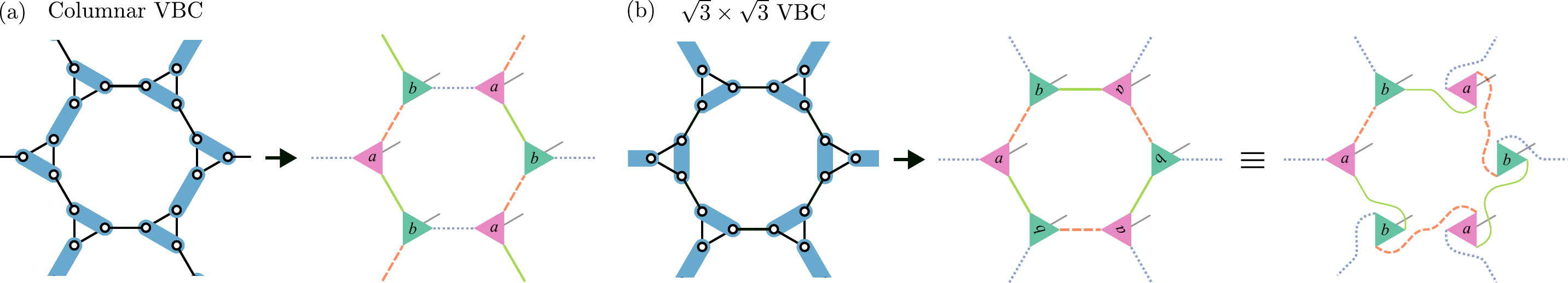}
     \caption{The two-trimer unit based setups that hosts only (a) the columnar VBC and (b) the $\sqrt{3}\times\sqrt{3}$ VBC state. Note that in the first setup, the two tensors, \( a \) and \( b \), are arranged in an $a$-$b$-$a$-$b$-$a$-$b$ sequence. In contrast, the second setup follows the ordering $a$-$b$-$R(a)$-$R(b)$-$R^2(a)$-$R^2(b)$ in counterclockwise direction, where \( R \) represents a rotation by \( 2\pi/3 \) about the center of the tensors.}
 \label{fig-two-tensor}
\end{figure*}

To obtain the GS, or at least a close approximation of it, the tensors $a_{i,j,k,s}^{[\mathbf{x}]}$ must be optimized such that the energy of the system is minimized. Various optimization strategies have been proposed in the literature, including the simple update~\cite{Jiang2008}, full update~\cite{Jordan2008,Phien2015}, and gradient-based methods~\cite{Hasik2021}. In this work, we employ the simple update algorithm, where we evolve a random initial tensor network in imaginary-time through a sequence of local, quasi-adiabatic updates, mimicking a slow annealing process that relaxes the system towards the ground state. For gapped systems, like ours, where entanglement is relatively short-ranged and correlations decay quickly, the simple update provides an efficient and sufficiently accurate approximation of the GS as the global environment doesn't drastically affect local updates. The imaginary-time evolution is performed by applying iteratively the Trotter-Suzuki decomposition of $e^{-\tau H}$ as follows
\begin{align}
   | \psi_{GS} \rangle  & = \lim_{\beta \rightarrow \infty} e^{-\beta H} | \psi_0 \rangle \\
  & \simeq \lim_{n \rightarrow \infty} \left(\prod_{ e \in \langle i, j \rangle } e^{- \tau H_{e}}\right)^n | \psi_0 \rangle \ .
\end{align}
In practice, we use $\tau = 10^{-3}$-$10^{-2}$. The imaginary-time evolution is then reduced to the contraction of a 3D tensor network, where two-site gates are successively applied to nearest-neighbor tensors. If left unchecked, the dimension of the local tensors $a^{[\mathbf{x}]}$ would grow exponentially fast with the number of gates operated on the network. After inserting each gate, the local tensors $a^{[\mathbf{x}]}$ are thus projected back to a relevant subspace of dimension $D$ by performing a singular value decomposition and keeping only the $D$-th largest singular values. 

Once the local tensors $a^{[\mathbf{x}]}$ have been optimized, local observables are measured by contracting the infinite 2D tensor network defined with local tensor \mbox{$A_{i^{},i',j^{},j',k^{},k'}^{[\mathbf{x}]} = a_{i',j',k',s}^{[\mathbf{x}]\dagger} a_{i^{},j^{},k^{},s^{}}^{[\mathbf{x}]}$} [see Fig.~\ref{fig-TN} (b)]. To do so, we use a variant of the corner transfer matrix renormalisation group (CTMRG) algorithm specifically designed to contract tensor networks on the honeycomb lattice \cite{gendiar2012,nyckees2023,lukin2023,lukin2024}. Initially introduced to compute partition functions in the context of statistical physics \cite{Nishino1996}, the CTMRG algorithm is nowadays mostly used to contract iPEPS wavefunctions \cite{Orus2009}. It approximates the infinite contraction by an environment $E^{[\mathbf{x}]}$ comprised of row $T^{[\mathbf{x}]}$ and corner tensors $C^{[\mathbf{x}]}$ [see Fig.~\ref{fig-TN} (b)], of dimension $\chi \times D^2 \times \chi$ and $\chi \times \chi$, respectively. The parameter $\chi$ plays the role of the control parameter, and one expects the contraction to converge to the exact results in the $\chi \rightarrow \infty$ limit. 
It is worth noting that the results become truly variational only in that limit. 
It is thus necessary to systematically make sure that the observables, here the energy, have converged in $\chi$. In our investigation of the SLHA, we use $\chi$ up to $D(D+2)+16$ for which the energy has sufficiently converged (with precision of $10^{-6}$) for all $D\in[2,11]$.

In our initial estimation of the GS of the SLHA, we consider a unit cell composed of six trimers as seen in Fig.~\ref{fig-TN} (a). In the small $J_d/J_t$ limit, the algorithm converges, depending on the choice of initial tensors, towards either the $\sqrt{3}\times\sqrt{3}$ VBC [Fig.~\ref{fig-vbs} (c)] or the columnar state [Fig.~\ref{fig-vbs} (d)] with nearly degenerate energies (difference $\sim 10^{-5}$). To disentangle the competition between these states, we subsequently introduce two alternative setups based on two-trimer unit cells, each designed to preserve only the symmetries compatible with either the $\sqrt{3}\times\sqrt{3}$ VBC or the columnar state. We illustrated the different setups in Fig.~\ref{fig-two-tensor}. The two-tensor setups also improve the convergence of the algorithm.

\begin{figure}[b]
    \includegraphics[width=0.8\columnwidth]{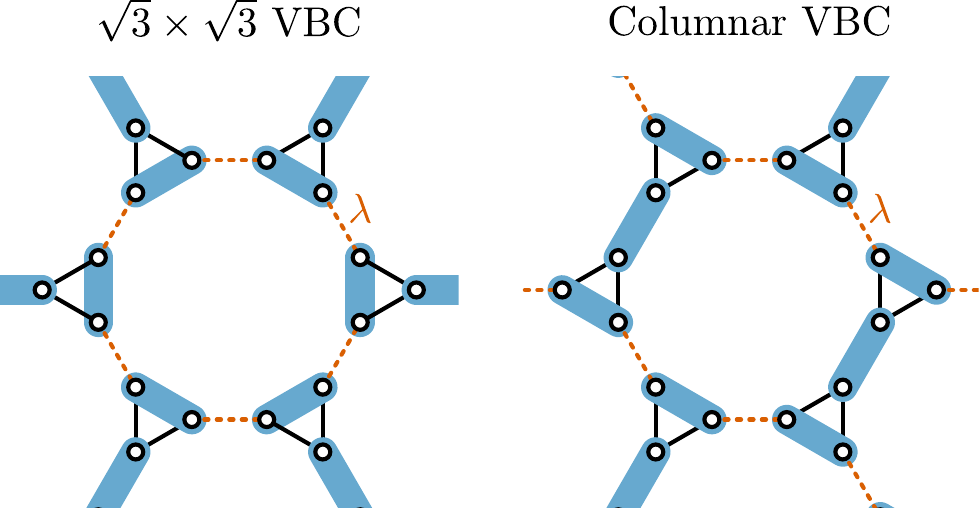}
    \caption{The deformation of the Hamiltonian~\eqref{eq-hamil} for the \mbox{$\sqrt{3}\times\sqrt{3}$} and columnar VBC states into bowties with the usual $J_t$ and $J_d$ couplings and the couplings between the bowties, $\lambda$, as a small perturbations}
    \label{fig-perturbation_setup}
\end{figure}

\subsection{Pertubative Analysis}
\label{sec:PerturbativeAnalysis}
To complement the iPEPS calculations, mainly for small ratios of $J_d/J_t$, we adopt a perturbative series expansions approach to understand the energy splitting between the $\sqrt{3}\times\sqrt3$ and columnar VBC.
Regarding~\eqref{eq-hamil}, a perturbative analysis in the limit of large $J_d$ is fairly straightforward due to the uniqueness of the product singlet ground state for $J_t=0$.
In opposition to that, the problem in the large $J_t$ limit is a degenerate one.
The ground-state degeneracy of isolated triangles $J_d=0$ grows exponentially with four to the power the number of triangles.
To get a perturbative grip on the energy of the states, we choose to deform the Hamiltonian~\eqref{eq-hamil} into bowties with the usual $J_t$ and $J_d$ couplings, and the couplings between the bowties as $\lambda$, which acts as a small perturbation parameter (refer to Fig.~\ref{fig-perturbation_setup}).
We then recover the original Hamiltonian for $\lambda=J_d$.
Note that each of the bowties has a unique GS where three singlets are placed: one on the dimer bond connecting the two triangles and two on the two bonds on the triangles that are not connected to the dimer.

The perturbative treatment of the inter-bowtie couplings is a non-degenerate problem where the bowties can be seen as supersites placed on a kagome (square) lattice for the $\sqrt3\times\sqrt3$ (columnar) VBC.
The local Hilbert space of each supersite has dimension $2^6=64$.
The remaining step to set up a perturbative expansion is to express the inter-bowtie coupling in the bowtie eigenbasis.
We do this numerically by diagonalizing the bowtie for given values of $J_t$ and $J_d$ and expressing the action of the coupling in this basis.
A perturbative expansion of a non-degenerate state can be efficiently done using the L\"owdin partitioning formalism \cite{Loewdin1962,Yao2000,Kalis2012}.
Within this method, the naive approach to obtain results in the thermodynamic limit is to evaluate the required operator sequences on a periodic cluster of supersites large enough such that no process reaches the periodicity of the cluster.
However, due to the large local Hilbert space dimension of the supersites, this approach is unfeasible  in practice.
For cluster-additive quantities~\cite{Hoermann2023}, like the ground-state energies, we are interested in, one can evaluate the energy density by linked-cluster expansion using a full graph-decomposition \cite{Gelfand2000,Oitmaa2006,Muehlhauser2024,Adelhardt2024}\footnote{As we have a numerical code at hand, we use a hypergraph framework to calculate the full graph decomposition \cite{Muehlhauser2022}. This is an unnecessary conceptional overhead that leads to the same final result.}. 
For the technical evaluation of the perturbative formalism on the graphs, we use the bookkeeping scheme described in \cite{Coester2015}.
Technically, it is important to fix an enumeration of the sites on a bowtie at the beginning of the calculation and keep track of the connections throughout the graph generation and perturbative calculation.
With the procedure described above, we were able to evaluate the perturbative series for the energy of both states up to order seven in $\lambda$.
We provide the bare series in the App.~\ref{app-GSSeries}.

We use the standard method of Pad\'e approximants to extrapolate the series beyond their original radius of convergence \cite{Guttmann1989}.
Given the series for the energy density in order $o_{\max}$, i.e.,
\begin{align}
\label{eq-GeneralSeries}
e_g(\lambda)=\sum_{i=0}^{o_{\max}} \epsilon_i\lambda^i=\epsilon_0+\epsilon_1\lambda+\epsilon_1\lambda^2+...+\epsilon_{o_{\max}}\lambda^{o_{\max}}
\end{align}
with coefficients $\epsilon_i\in\mathbb{R}$ from the perturbative expansion and $\lambda\in \mathbb{R}$ the perturbation parameter,
we define the Pad\'e approximant $G^{L/M}(\lambda)$ to be a fraction of two polynomials that give the series \eqref{eq-GeneralSeries} as a Taylor expansion in order $o_{\max}$ about $\lambda=0$
\begin{align}
    G^{L/M}(\lambda)=\frac{p_0+p_1\lambda+p_2\lambda^2+...+p_L\lambda^L}{q_0+q_1\lambda+q_2\lambda^2+...+q_M\lambda^M} \ .
\end{align}
The polynomial in the denominator (numerator) has degree $M$ ($L$).
The definition via the Taylor expansion provides a sufficient set of linear equations that can be solved for given parameters.
Usually, extrapolations with a small relative difference $|L-M|$ provide the best results.
In our analysis, extrapolations with unphysical singularities are sorted out.
Further, also defective Pad\'e approximants that have a singularity at the same point in the numerator and denominator are neglected.
We only present the leading extrapolations with $|L-M|\leq 1$; the other extrapolations show either a larger deviation to the iPEPS energy or have defective/unphysical behavior.

\begin{figure}[t]
    \includegraphics[width=0.95\columnwidth]{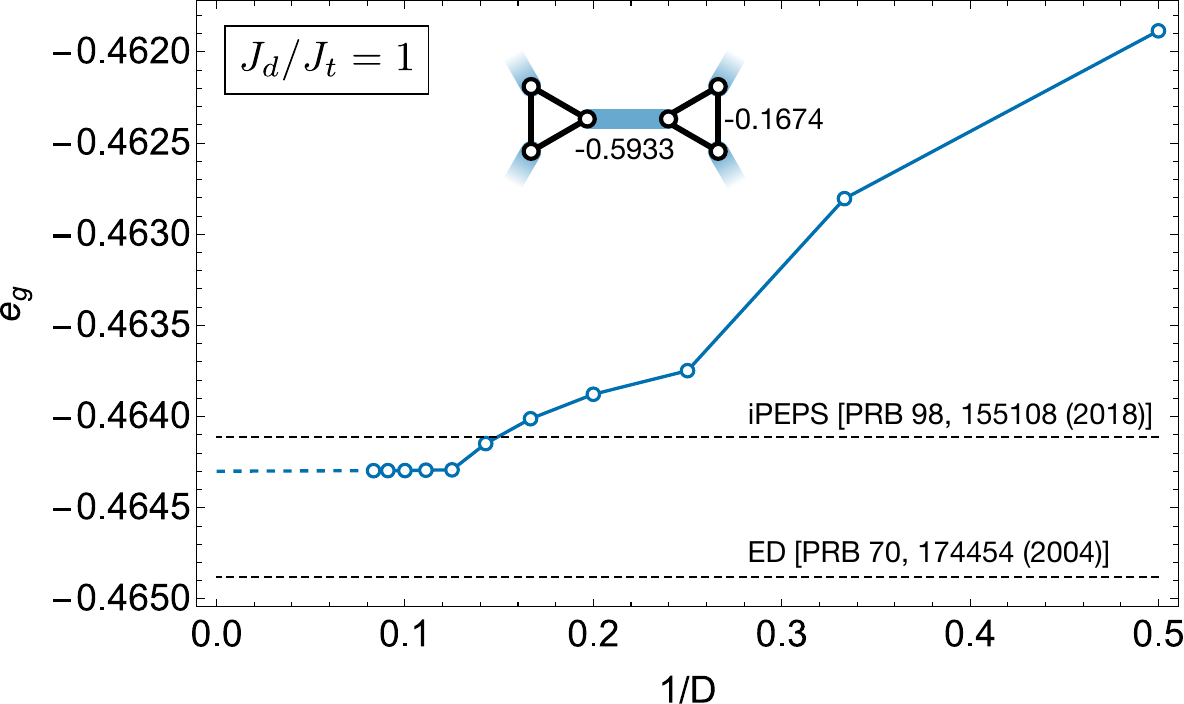}
    \caption{Scaling of the GS energies per site, $e_g$, obtained in iPEPS as a function of the bond dimension $D$ at the isotropic point, $J_d = J_t = 1$. The spin-spin correlations on the two types of bonds are also shown. The GS is a uniform dimer VBS state. The excellent energy convergence for the larger values of $D$ is due to the low entanglement of the GS.} \label{fig-isotropic}
\end{figure}

\section{Results}\label{sec:results}
For $S=1/2$, it is apparent that in the limiting case of $J_t\to0$ the GS is a product state of isolated singlets on the $J_d$ bonds. By introducing the $J_t$ couplings, one induces frustration into the system. However, these couplings only renormalize the state in second order of the perturbation resulting in a stable dimer VBS state for weak enough $J_t$. This VBS state has a six-site unit-cell and fully respects the symmetries of the SLHA Hamiltonian \eqref{eq-hamil}. 
At $J_t / J_d = 0.05$, using the iPEPS approach, we obtain a ground-state energy of $-0.375234$ in the $D \to \infty$ limit. This result is in excellent agreement with the previous iPEPS and DMRG studies~\cite{Star-iPEPS1,Star-iPEPS2} and perfectly matches with the second-order perturbative energy
\begin{align*}
e_0 = -\frac{3}{8} J_d -\frac{3}{32} \frac{J_t^2}{J_d}
\end{align*}
obtained from the non-degenerate perturbative expansion of the dimer VBS in $J_t/J_d$.

We find that this VBS state continues to be the GS as the relative strength of $J_d/J_t$  decreases. For the isotropic model, i.e., $(J_d,J_t)=(1,1)$, we find it in our iPEPS calculations to be stable with an energy of $-0.464299$ in the $D\to\infty$ limit (see Fig.~\ref{fig-isotropic}). The VBS state has bond energies $-0.5933$ on the $J_d$ bonds and $-0.1674$ on the $J_t$ bonds. This is in accord with previous studies using iPEPS, DMRG, and Gutzwiller projected states~\cite{Star-ED,Star-iPEPS1,Star-iPEPS2,Star_PSG}. For this case, we achieve excellent energy convergence for all the values of $D\ge8$ signaling prevailing local entanglements. 

\begin{figure}[t]
    \includegraphics[width=0.9\columnwidth]{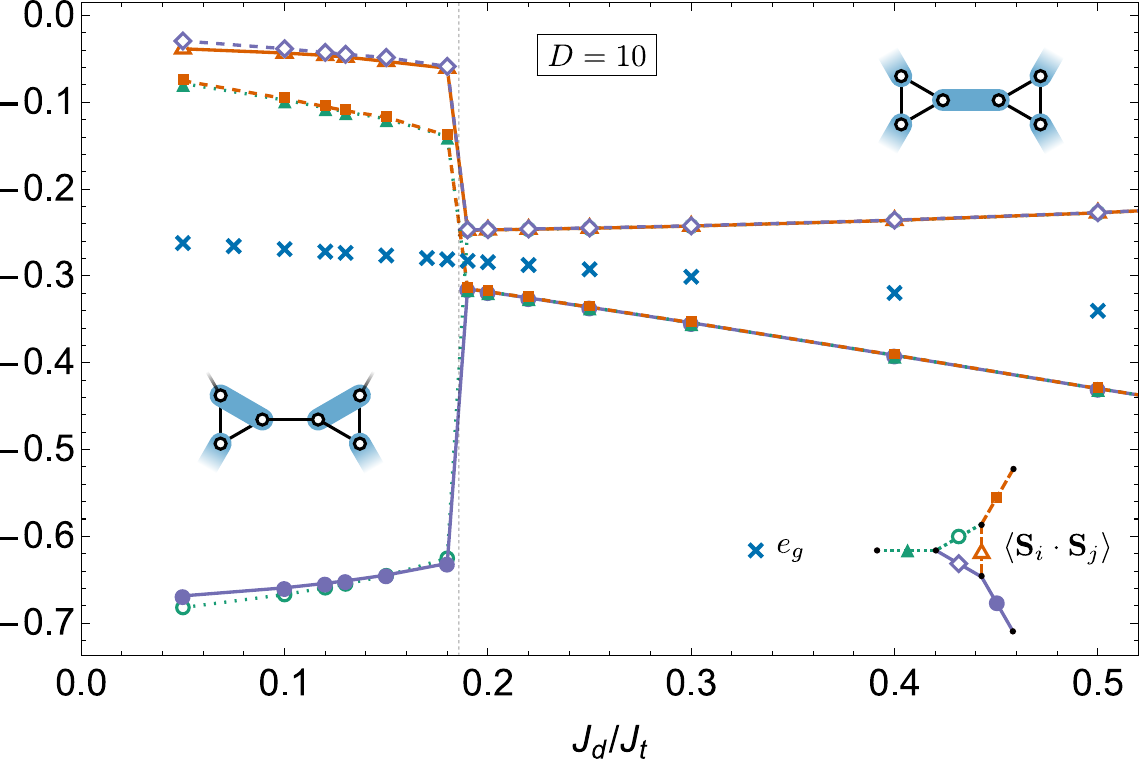}
    \caption{GS energies per spin, $e_g$, and the spin-spin correlations, $\braket{\mathbf{S}_i\cdot \mathbf{S}_j}$, on a trimer and its surrounding dimer bonds with respect to $J_d/J_t$ for fixed iPEPS bond dimension, $D=10$. The spin-spin correlations on the $J_t$ ($J_d$) bonds are shown by open (solid) markers. For \( J_d / J_t > 0.18 \), spin-spin correlations are uniform within both trimer and dimer bonds, with stronger AFM correlations on the dimer bonds, consistent with the uniform dimer VBS state. The results show a first-order transition out of the dimer VBS phase at $J_d/J_t=0.185(5)$, into the $\sqrt{3}\times\sqrt{3}$ VBC phase indicated by the development of strong AFM correlations on one of the trimer bonds and the dimer bond orthogonal to it. The different lines are only guide for the eye.} \label{fig-corr}
\end{figure}

Beyond the isotropic point, one would naively expect that the stronger frustrations will destabilize the dimer VBS and the system will undergo a phase transition. 
In contrast, we see a transition out of the dimer VBS phase occurring only at a smaller ratio of $J_d/J_t\sim 0.185(5)$ for all $D\ge 5$.
This is illustrated in Fig.~\ref{fig-corr} where we show the GS energy as well as the spin-spin correlations on the $J_t$ and $J_d$ bonds for $D=10$. Although it is difficult to identify the nature of the phase transition from the GS energy alone, the spin-spin correlations clearly indicate that the transition is first-order.

\begin{figure}[t]
    \includegraphics[width=0.95\columnwidth]{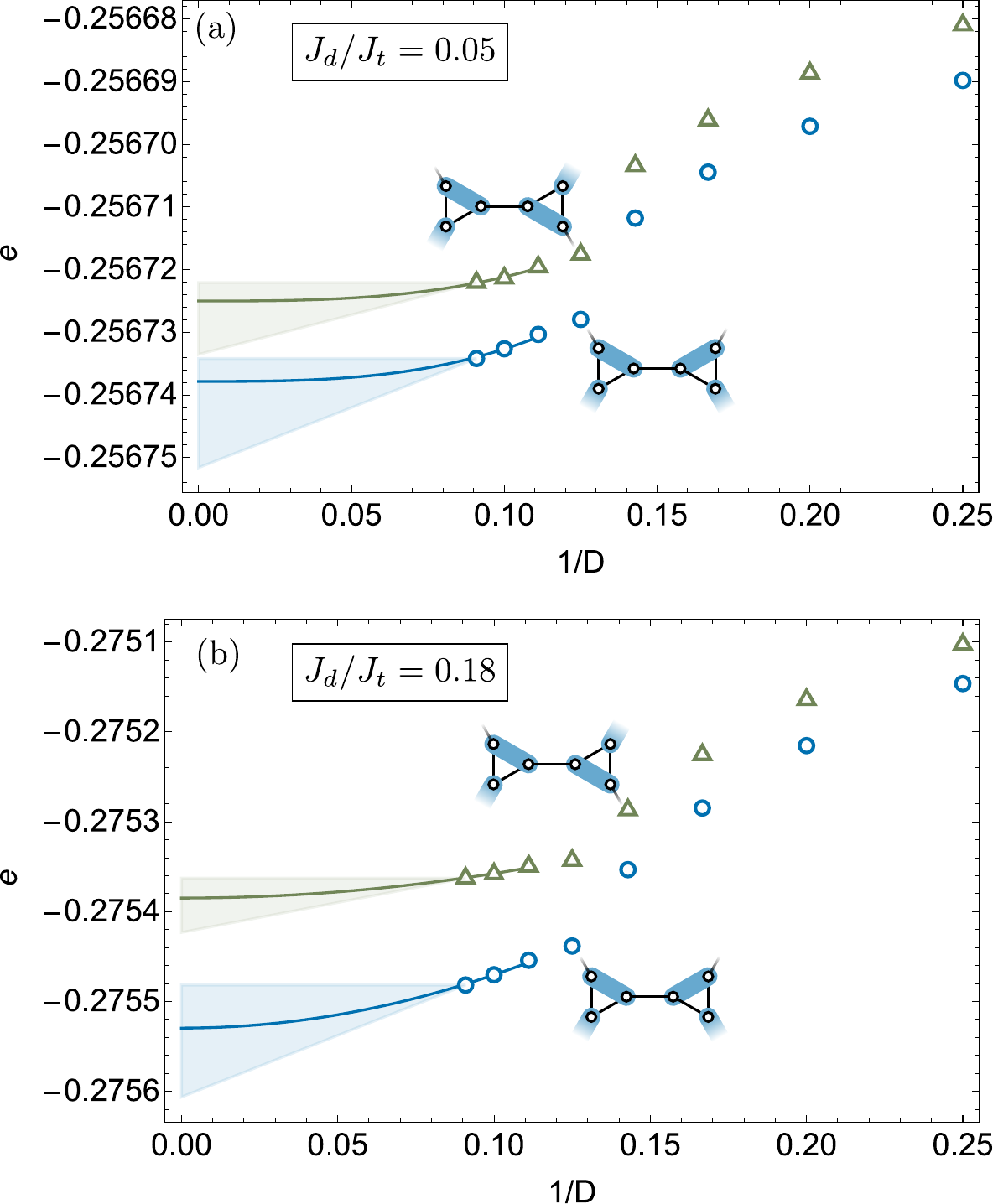}
    \caption{Scaling of the energies per spin for the $\sqrt{3}\times\sqrt{3}$ and the columnar VBC state obtained from the two-tensor iPEPS setup as a function of the bond dimension $D$ at (a) $J_d/J_t=0.05$ and (b) $J_d/J_t=0.18$. The solid lines shows the extrapolation for $D\to\infty$ obtained via an exponential fitting of the data for $D=8,9,10$ and $11$. The shaded region shows a conservative estimate of the error bars corresponding to the energies; the upper bounds are the energies for $D=11$ and the lower bound is obtained by a linear extrapolation of the energies of $D=10$ and $11$.} \label{fig-energy_small_Jd}
\end{figure}

Although the six-tensor setup allows us to clearly identify the phase transition, the precise nature of the resulting phase remains difficult to identify. Although the energy converges fairly quickly as a function of $D$ for $D\ge8$ in the dimer VBS phase, the energy for $J_d/J_t\le0.18$ still shows significant variation at the largest $D$ used in our iPEPS simulations. More importantly, there is a sizable number of low-lying states for $J_d\ll J_t$ and a number of these states can be hosted in the six-tensor setup. We find that upon a random initialization of the state, the system invariably assumes either one of the two distinct VBC states at zero temperature, namely the $\sqrt{3}\times\sqrt{3}$ VBC and the columnar VBC depicted in Fig.~\ref{fig-vbs} (c) and (d), respectively. 

To capture the nature of the ground state for $J_d/J_t\le0.18$, we switch from
using the six tensor setup to the two tensor setup. Note that both two-tensor setups can host the dimer VBS, but, where the setup in Fig.~\ref{fig-two-tensor} (a) can only bear the columnar VBC state, the other setup [in Fig.~\ref{fig-two-tensor} (b)] can only hold the $\sqrt{3}\times\sqrt{3}$ VBC. This allows us to carefully track the energies of the competing VBC states separately for all ratios of $J_d\ll J_t$. 
In Fig.~\ref{fig-energy_small_Jd}, we show the scaling of the energies of the two VBC states at two representative points, namely, $J_d/J_t=0.05$ and $J_d/J_t=0.18$.
We find that for $J_d/J_t\le0.18$, the energy corresponding to the $\sqrt{3}\times\sqrt{3}$ VBC is consistently lower than that of the columnar VBC for all finite $D$, as well as the extrapolations for $D\to\infty$.

 \begin{figure}
    \centering
    \includegraphics[width=0.95\linewidth]{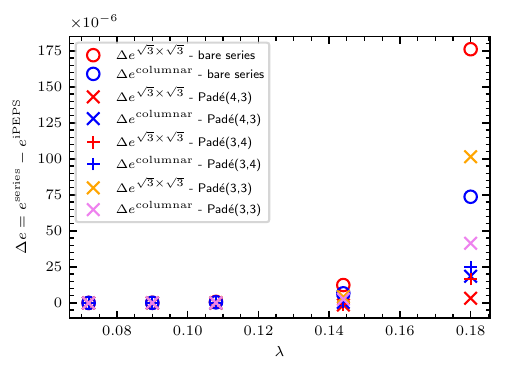}
    \caption{Comparison of the perturbative analysis for several $\lambda$ values for $J_d=0.18$ against non-perturbative iPEPS calculations using the two-tensor setup. The original undeformed SLHA corresponds to $\lambda=J_d$. The \( (3,3) \) Padé approximants are derived using terms from the series expansions up to order six.}
    \label{fig-convergence}
\end{figure}

To extend our understanding of the ground state in the challenging regime of small $J_d/J_t$, we employ the perturbative analysis discussed in Sec.~\ref{sec:PerturbativeAnalysis}.
With the deformation of the Hamiltonian into uncoupled bowties (see Fig.~\ref{fig-perturbation_setup}) and the expansion in the inter-bowtie couplings, we obtain a grasp on the energies of both VBC states by non-degenerate perturbation theory.
We obtain the energies as a power series up to order seven in the deformation parameter $\lambda$.

In order six in $\lambda$, the first difference between the two energy series occurs and the $\sqrt{3}\times\sqrt{3}$ VBC state becomes lower in energy and the splitting become more pronounced in the seventh order.
Note that the perturbative problem for the $\sqrt{3}\times\sqrt{3}$ (columnar) VBC has bowtie supersites coupled via $\lambda$-bonds on a kagome (square) lattice. 
Heuristically, one rule we observe for the full graph expansion problem at hand is that each bond on a graph needs to be acted on twice by the perturbation in order to contribute. The exception to this rule is the six-site ring graph around the hexagons of the effective kagome lattice.
The latter one corresponds to a dimerized chain in the original spin picture.
Therefore, the energy difference between the states can be attributed in the graph expansion by the possibility to embed a three-site ring graph on the kagome lattice and the ring of six around a hexagon.
Despite the full graph expansion, we cannot evaluate the perturbative expansion further than order seven due to memory and runtime limitations.

We also evaluate non-perturbative energy values for the two candidates for the deformed model for given values of $\lambda$ using iPEPS.
We use this to gauge the convergence of our series and the Pad\'e approximants.
We compare the results from both methods in Fig.~\ref{fig-convergence}.
We see that for small $\lambda$ values, both techniques are in perfect agreement, while approaching the undeformed case $\lambda=J_d$, the results begin to differ.
We find that the bare series expansion provides systematically larger estimates for the energy compared to the iPEPS calculations.
However, by comparing the sixth-order Pad\'e (3,3) extrapolant with the seventh-order Pad\'e extrapolants, one can see that with increasing order in $\lambda$ from six to seven, the series results tend to the iPEPS values.
We would like to highlight the consistency.

 \begin{figure}
    \centering
    \includegraphics[width=0.95\linewidth]{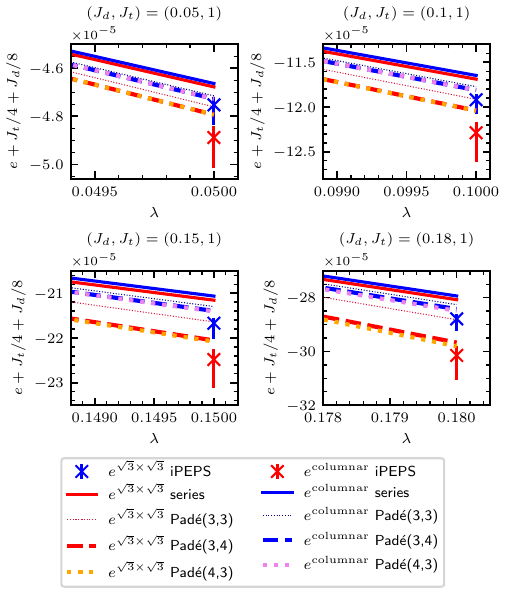}
    \caption{Energies for the $\sqrt{3}\times\sqrt{3}$ VBC state and the columnar VBC state around the undeformed case $\lambda=J_d$ obtained by the iPEPS two-tensor setup calculations and the perturbative analysis. The four panels display four representative values of $(J_d,J_t)$.}
    \label{fig-resultsoverview}
\end{figure}

Finally, we compare the estimates for energies in the undeformed case,  $\lambda=J_d$, from the two methods in Fig.~\ref{fig-resultsoverview}.
As discussed, there is an energy splitting between the two states found in both the iPEPS calculations and the perturbative treatments.
For all presented Padé approximants the $\sqrt{3}\times\sqrt{3}$ VBC are always energetically lower than the ones for the columnar VBC.
We also see the systematic shift of the extrapolated series towards higher energies and the lowering of the energies when going from the order six Pad\'e (3,3) to the order seven Pad\'es (3,4) and (4,3).
Note that, although the precise iPEPS energy values cannot be produced with the extrapolation of the series in order seven, the magnitude of the energies, as well as the energy splitting, is approaching those obtained from the iPEPS. 

\section{Conclusions and Discussions}\label{sec:conclusions}
We investigate the ground state (GS) phase diagram of the spin-$1/2$ Heisenberg antiferromagnet on the star lattice (SLHA). 
The lattice has two symmetry-inequivalent bonds which we assign with coupling constants $J_d$ and $J_t$ [see Fig.~\ref{fig-vbs} (a)]. The $J_t$ bonds make triangles, and the $J_d$ bonds link the $J_t$-triangles. 
From iPEPS calculations at the isotropic point \mbox{$(J_d,J_t)=(1,1)$}, we obtain a VBS state with strong singlets on all $J_d$ bonds, which fully respects the symmetries of the SLHA. 
Apart from this state, we found another valence bond state, a VBC, that appears in the GS phase diagram of \eqref{eq-hamil} for different Heisenberg coupling ratios. This VBC is a three-fold degenerate $\sqrt{3}\times\sqrt{3}$ valence bond order which breaks the lattice rotation symmetry and has a 18-site unit-cell. We have also identified another VBC state, deemed as columnar VBC state, which competes closely with the $\sqrt{3}\times\sqrt{3}$ VBC. Upon applying the perturbative analysis we calculated the energies of both the states as a power series up to seventh-order in a deformation parameter (see Fig.~\ref{fig-two-tensor}). We see that the $\sqrt{3}\times\sqrt{3}$ VBC state has a slightly lower energy in order six which splits even further in order seven.
The Pad\'e approximants of the perturbative series show excellent qualitative and quantitative agreement with the iPEPS results.

From iPEPS simulations, we find a first-order phase transition between these two phases at $J_d/J_t\approx0.18$ (for all $D\in[5,11]$), which is much smaller than $J_d/J_t\sim 1$ as discussed in previous literature. The higher estimate originates from the ED study~\cite{Star-ED}, which reports many low-energy states for $J_d/J_t\le0.77$. 
Despite the apparent discrepancy, a closer examination reveals that the ED results and our findings are not in conflict.
These low-energy states observed in ED are the family of resonating valence bond (RVB) states with singlet coverings that maximizes the number of occupied trimer bonds. One would naturally expect one of these states to eventually become the GS when $J_t$ is sufficiently strong. The momentum-resolved ED data identifies the \( \sqrt{3} \times \sqrt{3} \) VBC as a particularly low-lying excited state whose energy significantly drops as the system size increases. However, clusters up to 108 sites do not observe this state becoming the ground state in the calculated range of \( J_d / J_t \ge 0.33 \). Instead, the lowest-energy state throughout this range has zone-center momentum, suggesting it is in fact the dimer VBS.
In our investigation of dimer VBS for $J_d/J_t<1$, we observe that while the AFM spin-spin correlation on the dimer bonds remain stronger than that of the trimer bonds, the correlations on the dimer bonds reduce rapidly as $J_d$ decreases, whereas the correlations on the trimer bonds remain nearly unchanged. Their difference diminishes significantly as one approaches the phase transition (Fig.~\ref{fig-corr}). In this sense, the correlation profile of the dimer VBS state for small $J_d$ resembles more an RVB state than a typical VBS. This internal resonance within the $J_t$ bonds helps to stabilize the dimer VBS phase even at low $J_d/J_t$. Our findings thus align with the ED observation of the low-lying \( \sqrt{3} \times \sqrt{3} \) VBC state, but show that its emergence as the ground state occurs at a lower \( J_d / J_t \), and not immediately following the level collapse of the excited RVB states around \( J_d / J_t \approx 0.77 \).

The three states discussed in this article compete closely for small $J_d/J_t$, so even a small correction to the energies—arising from finite \( D \) effects—can induce a prominent shift in the crossing point. Here, the rapid energy convergence of the dimer VBS for \( D \ge 8 \) contrasts with the behavior of the symmetry-broken states, whose energies continue to vary significantly at the largest \( D \). Rather than identifying the phase transition via energy comparison, we rely on the iPEPS's ability to spontaneously assume a symmetry-broken state during imaginary time evolution. The symmetry-broken state appears at $J_d/J_t\le 0.18$ for all $D\in[5,11]$. 
The current iPEPS algorithm developed for the star lattice, and in particular the honeycomb CTMRG, is well-suited to study such symmetry breakings as it preserves the lattice’s symmetry.  
Among various possibilities, the non-physical symmetry breaking observed in previous iPEPS study~\cite{Jahromi2018} may arise from the contraction of the infinite tensor network using the square CTMRG algorithm. The honeycomb CTMRG offers a robust tool to explore models with $C_3$ symmetry such as the star lattice, kagome, or the maple-leaf lattice.

Our findings and the identified phases may have direct relevance to experimental realizations of star lattice magnets, such as polymeric iron acetate $\mathrm{Fe_3(\mu_3\textrm{-}O)( \mu \textrm{-}OAc)_6(H_2O)_3][Fe_3( \mu_3\textrm{-}O)( \mu \textrm{-}OAc)_{7.5}]_2\cdot}$ $\mathrm{7H_2O}$~\cite{Star_Fe} and hybrid copper sulfate $\mathrm{[(CH_3)_2(NH_2)]_3Cu_3(OH)(SO_4)_4\cdot0.24H_2O}$~\cite{Star_Cu}. While iron acetates ultimately exhibit magnetic ordering at 4.5~K~\cite{Star_Fe}, hybrid copper sulfates resist magnetic ordering down to 0.1~K~\cite{Star_Cu, PhysRevB.109.L180401}. The latter one has an estimated coupling ratio $J_d/J_t\approx 0.06$, placing it deep within the regime where we find the \( \sqrt{3} \times \sqrt{3} \) VBC. Moreover, the sizable Dzyaloshinskii–Moriya (DM) interaction in this compound~\cite{PhysRevB.109.L180401} could play a crucial role in the stability of the valence bond phases, further enriching the underlying physics~\cite{esaki2025spinnernstthermalhall}. Additionally, the study the thermal phase transitions corresponding to the three-fold degenerate $\sqrt{3}\times\sqrt{3}$ VBC state is also an interesting avenue to pursue~\cite{Star-Thermal}. We defer the exploration of these effects to future studies.

\textit{Acknowledgments:}
The work at EPFL is supported by the Swiss National Science Foundation Grant No. 212082.
Numerical computations performed at EPFL have been done using the facilities of the Scientific IT and Application Support Center of EPFL (SCITAS).
The work at FAU was funded by the Deutsche Forschungsgemeinschaft (DFG, German Research Foundation) - Project-ID 429529648 - TRR 306 QuCoLiMa (Quantum Cooperativity of Light and Matter).
KPS and JK acknowledge the support by the Munich Quantum Valley, which is supported by the Bavarian state government with funds from the
Hightech Agenda Bayern Plus. 
PG and JK thanks  P. Adelhardt for useful discussions.

\bibliography{Refs.bib}

\appendix
\section{Energy series}
\label{app-GSSeries}

\subsection{Series for $J_d=0.05$ and $J_t=1$}

\begin{align*}
		e_{J_d=0.05}^{\sqrt{3}\times \sqrt{3}} = &-0.256250000000000-0.142137096774194\lambda^2\\
        &-0.439654331425602\lambda^3-4.9319087793544\lambda^4\\
        &-43.4187580105247\lambda^5-510.772757663832\lambda^6 \\
		&-6515.72306154855\lambda^7
\end{align*}

\begin{align*}
		e_{J_d=0.05}^{\text{columnar}} = 
        &-0.256250000000000-0.142137096774194\lambda^2\\
        &-0.439654331425603\lambda^3-4.93190877935443\lambda^4\\
        &-43.4187580105255\lambda^5-474.73472763576\lambda^6\\
		&-5422.46656667045\lambda^7
\end{align*}

\subsection{Series for $J_d=0.10$ and $J_t=1$}

\begin{align*}
		e_{J_d=0.10}^{\sqrt{3}\times \sqrt{3}} =
        & -0.262500000000000-0.0898437499999978\lambda^2\\
        &-0.138580322265618\lambda^3-0.690449120502133\lambda^4\\
        &-3.14158093119301\lambda^5-18.8032247915941\lambda^6\\
	    &-126.418077170417\lambda^7
\end{align*}

\begin{align*}
		e_{J_d=0.10}^{\text{columnar}} =
        & -0.262500000000000-0.0898437499999979\lambda^2\\
        &-0.138580322265618\lambda^3-0.690449120502132\lambda^4\\
        &-3.141580931193\lambda^5-17.22003445278\lambda^6 \\
		&-101.071090052366\lambda^7
\end{align*}

\subsection{Series for $J_d=0.15$ and $J_t=1$}

\begin{align*}
		e_{J_d=0.15}^{\sqrt{3}\times \sqrt{3}}=
        &-0.268750000000000-0.0722853535353534\lambda^2\\
        &-0.0767149079175582\lambda^3-0.231728170574219\lambda^4\\
        &-0.725227445636052\lambda^5-2.9742094792858\lambda^6\\
		&-14.1215391619514\lambda^7
\end{align*}

\begin{align*}
		e_{J_d=0.15}^{\text{columnar}} 
        &= -0.268750000000000-0.0722853535353534\lambda^2\\
        &-0.0767149079175579\lambda^3-0.231728170574217\lambda^4\\
        &-0.725227445636026\lambda^5-2.67695026976006\lambda^6\\
		&-10.7786901340191\lambda^7
\end{align*}

\subsection{Series for $J_d=0.18$ and $J_t=1$}

\begin{align*}
		e_{J_d=0.18}^{\sqrt{3}\times \sqrt{3}}=
        &-0.272500000000000-0.0663855820105819\lambda^2\\
        &-0.0603480822949036\lambda^3-0.145139523887787\lambda^4\\
        &-0.385639794276134\lambda^5-1.34533169065084\lambda^6 \\
	    &-5.52002807294595\lambda^7
\end{align*}

\begin{align*}
		e_{J_d=0.18}^{\text{columnar}} = 
        &-0.272500000000000-0.0663855820105819\lambda^2\\
        &-0.0603480822949036\lambda^3-0.145139523887786\lambda^4\\
        &-0.385639794276142\lambda^5-1.19729560499442\lambda^6 \\
		&-4.08895533616881\lambda^7
\end{align*}

\end{document}